\documentclass[fleqn,usenatbib,useAMS]{mnras}

\usepackage{graphicx}	
\usepackage{amsmath}	
\usepackage{amssymb}	
\usepackage{csquotes}
\usepackage{lmodern}    
\usepackage{multicol}   
\usepackage{bm}		    
\usepackage{pdflscape}	
\usepackage{xcolor}

\usepackage[T1]{fontenc}
\usepackage{ae,aecompl}
\usepackage{newtxtext,newtxmath}

\title[NIHAO XXVI: Nature versus nurture]{NIHAO XXVI: Nature versus nurture, the Star Formation Main Sequence and the origin of its scatter}
\author[M. Blank, L. E. Meier, A.V. Macci\`o, A. A. Dutton, K. L. Dixon, N. H. Soliman, X. Kang]{
\parbox{\paperwidth}{
Marvin Blank,$^{1,2,3}$\thanks{marvin.blank@nyu.edu}
Liam E. Meier,$^{1}$
Andrea V. Macci\`o,$^{1,2,4}$
Aaron A. Dutton,$^{1}$ \\
Keri L. Dixon,$^{1,2}$
Nadine H. Soliman$^{1,2}$
and Xi Kang$^{5,6}$\\}\\
$^{1}$New York University Abu Dhabi, PO Box 129188, Saadiyat Island, Abu Dhabi, United Arab Emirates\\
$^{2}$Center for Astro, Particle and Planetary Physics (CAP$^3$), New York University Abu Dhabi \\
$^{3}$Institut f\"{u}r Theoretische Physik und Astrophysik, Christian-Albrechts-Universit\"{a}t zu Kiel, Leibnizstr. 15, D-24118 Kiel, Germany\\
$^{4}$Max Planck Institut f\"{u}r Astronomie, K\"{o}nigstuhl 17, D-69117 Heidelberg, Germany\\
$^{5}$Zhejiang University-Purple Mountain Observatory Joint Research Center for Astronomy, Zhejiang University, Hangzhou 310027, China\\
$^{6}$Purple Mountain Observatory, No 8 Yuanhua Road, Nanjing 210034, China\\}

\date{\today}

\pubyear{2019}

\begin{document}
\label{firstpage}
\pagerange{\pageref{firstpage}--\pageref{lastpage}}
\maketitle

\begin{abstract}
We investigate how the NIHAO galaxies match the observed star formation main sequence (SFMS) and what the origin of its scatter is.
The NIHAO galaxies reproduce the SFMS and generally agree with observations, but the slope is about unity and thus significantly larger than observed values.
This is because observed galaxies at large stellar masses, although still being part of the SFMS, are already influenced by quenching.
This partial suppression of star formation by AGN feedback leads to lower star formation rates and therefore to lower observed slopes.
We confirm that including the effects of AGN in our galaxies leads to slopes in agreement with observations. 
We find the deviation of a galaxy from the SFMS is correlated with its $z=0$ dark matter halo concentration and thus with its halo formation time. This means galaxies with a higher-than-average star formation rate (SFR) form later and vice versa.
We explain this apparent correlation with the SFR by re-interpreting galaxies that lie above the SFMS (higher-than-average SFR) as lying to the left of the SFMS (lower-than-average stellar mass) and vice versa. Thus later forming haloes have a lower-than-average stellar mass, this is simply because they have had less-than-average time to form stars, and vice versa.
It is thus the nature, i.e. how and when these galaxies form, that
sets the path of a galaxy in the SFR versus stellar mass plane.

\end{abstract}


\section{Introduction}

It is now well established that most star-forming galaxies lie on the star formation main sequence (SFMS), which is a linear relation (in log-space) between the galaxy's star formation rate (SFR) and stellar mass 
\citep{2004_Brinchmann_Charlot_White,2007_Noeske_Weiner_Faber}.
The existence of the SFMS has been observed at $z=0$ \citep{2015_Renzini_Peng,2011_Elbaz_Dickinson_Hwang}, but also at larger redshifts \citep{2007_Daddi_Dickinson_Morrison, 2007_Elbaz_Daddi_LeBorgne,2015_Schreiber_Pannella_Elbaz,2015_Tasca_LeFevre_Hathi}.
The slope of the SFMS usually has values of 0.5-1, and shows no significant evolution with redshift, while its normalization is increasing with redshift.
The scatter of the SFMS is uniform with 0.2-0.3 dex for all observed redshifts.
\citet{2014_Speagle_Steinhardt_Capak} compile data from 25 different publications and study the evolution of the SFMS, they find that 0.1 dex of the scatter is due to the usage of different techniques among the different studies (\lq inter publication scatter\rq), after correcting for this effect the intrinsic scatter is 0.2 dex.

\citet{2015_Sparre_Hayward_Springel} investigate the SFMS with the Illustris simulations and reproduce the SFMS at $z=0$ and 4, but for intermediate redshifts of $z=1$ and 2 the normalization is slightly lower than observed values. The scatter is about 0.2-0.3 dex and thus consistent with observations.
The IllustrisTNG simulations \citep{2019_Donnari_Pillepich_Nelson} reproduce the SFMS at $z=0$ with 0.3 dex scatter, the slope and normalization are consistent with observations. Their scatter is constant with stellar mass and decreasing with redshift, their normalization is systematically lower than observations at $z \sim 0.75-2$ by 0.2-0.5 dex.
Also hydrodynamic simulation from \citet{2014_Kannan_Stinson_Maccio} and \citet{2014_Torrey_Vogelsberber_Genel} produce a tight relation between SFR and stellar mass.

\citet{2019_Matthee_Schaye} investigate the scatter of the SFMS and find that galaxies that are located above the SFMS tend to stay there for timescales of about 10 Gyr, thus it seems that \enquote{a galaxy's SFR remembers its past SFR.} They also report that later forming haloes tend to host galaxies with a higher SFR, and that the scatter of the SFMS correlates with the dark matter halo formation time.
\citet{2010_Dutton_Bosch_Frank} use a semi analytical model to investigate the scatter of the SFMS. They find that the scatter is mostly caused by variations in the galaxy's gas accretion history, and is moreover dependent on the halo concentration.

In this paper we investigate the SFMS with the NIHAO galaxies. These provide a higher spatial resolution and extend to lower stellar masses than earlier studies.
We show that the slope of the SFMS at higher stellar masses ($\gtrsim 10^{9}\,\mathrm{M}_{\odot}$) is already affected by AGN feedback, leading earlier works to predict a slope lower than unity.
We show that by incorporating smaller stellar masses and considering galaxies unaffected by AGN feedback, the slope of the SFMS is approximately
unity, and thus larger than predicted by earlier works.
We also investigate the origin of the scatter of the SFMS.
We confirm findings by \citet{2019_Matthee_Schaye} that the scatter of the SFMS is correlated with the halo formation time: haloes that form later have a higher than average SFR and vice versa.
We provide a new interpretation for this correlation: Galaxies with a higher-than-average SFR (located above the SFMS) can be re-interpreted as galaxies with a lower-than-average stellar mass (located to the left of the SFMS). Thus later (earlier) forming galaxies have a lower (higher) than-average stellar mass, which is simply because they have had less (more) time to form stars than earlier (later) forming galaxies.

The outline of this work is as follows:
In Section \ref{sec:nihao} we introduce the NIHAO simulations and describe how we measure the quantities used in this paper.
In Section \ref{sec:results1} we calculate the SFMS of the NIHAO galaxies, its slope, normalization and scatter, and compare our results with several observations.
In Section \ref{sec:results2} we investigate how and why galaxies deviate from the SFMS, i.e. what is causing the scatter in the relation between SFR and stellar mass. In Section \ref{sec:summary} we summarize our findings.

\section{The NIHAO galaxies}\label{sec:nihao}

We use the NIHAO suite of galaxy simulations, which consists of more than 150 zoom-in simulations of galaxies introduced by \citet{2015_Wang_Dutton_Stinson} and extended by \citet{2019_Blank_Maccio_Dutton}. The initial conditions originate from cosmological simulations with box sizes of 60, 20 and 15 $\rmn{Mpc} \, \rmn{h}^{-1}$ and with $400^3$ particles that are evolved until redshift zero, then haloes from these boxes are selected and resimulated individually with a higher resolution and with gas particles.
The resolutions of the zoom-in simulations are chosen such that each galaxy is resolved with about $10^6$ particles and we resolve the mass profile at $\leq$ 1 per cent of the virial radius. Thus we reach dark matter particle masses of $3.4 \times 10^3$ to $1.7 \times 10^6 \, \rmn{M}_{\sun}$ and dark matter softening lengths of 116 to 931 pc.
The ratio of dark and gas particle mass equals the cosmological mass ratio of dark matter and baryons of $\Omega_{\rmn{DM}}/\Omega_{\rmn{b}} = 5.48$,
the ratio of dark and gas particle softening length equals  $(\Omega_{\rmn{DM}}/\Omega_{\rmn{b}})^{1/2} = 2.34$.
We use a flat LCDM cosmology with parameters from the \citet{2014_Planck_Collaboration}.

NIHAO is simulated with an updated version \citep{2014_Keller_Wadsley_Benincasa} of the TreeSPH code {\sc Gasoline2} \citep{2017_Wadsley_Keller_Quinn}.
For gas cooling we consider hydrogen, helium, and various metal-lines in a uniform ultraviolet ionizing background \citep{2010_Shen_Wadsley_Stinson}, including photoionization, UV background heating \citep{2012_Haardt_Madau}, and Compton cooling.
Star formation occurs for gas particles that surpass a density and temperature threshold ($T < 15000 \, \rmn{K}$, $n > 10.3 \, \rmn{cm}^{-3}$) with a rate of $\dot{M_{\star}} = c_{\star} M_{\rmn{gas}} t_{\rmn{dyn}}^{-1}$, where $t_{\rmn{dyn}} = (4 \uppi G \rho)^{-1/2}$ is the gas particle's dynamical time, $\rho$ its density, $M_{\rmn{gas}}$ its mass and $c_{\star}=0.1$ the star formation efficiency.
To model supernova feedback we use the blastwave formalism of \citet{2006_Stinson_Seth_Katz}, where star particles with $8 < M_{\star}/M_{\sun} < 40$ inject metals and thermal energy to surrounding gas particles 4 Myr after they have formed, subsequently the cooling of these gas particles is delayed for $\sim 30 \, \rmn{Myr}$.
Following \citet{2013_Stinson_Brook_Maccio} star particles provide \lq early stellar feedback\rq \, before they produce a supernova, where 13 per cent of the total stellar flux of $2 \times 10^{50} \, \rmn{erg} \, \rmn{M}_{\sun}^{-1}$ is injected into the surrounding gas as thermal energy.
The free parameters of the stellar and supernova feedback model have been adjusted to match the $M_{\star}$-$M_{{200}}$ relation for one Milky Way-like galaxy at $z=0$.
For more details on the NIHAO project see \citet{2015_Wang_Dutton_Stinson}.

NIHAO has been very successful in reproducing galaxy properties for halo masses of $M_{200} \leq 2 \times 10^{12}\,\rmn{M}_{\sun}$, e.g., the stellar mass versus halo mass relation \citep{2015_Wang_Dutton_Stinson}, the galaxy velocity function \citep{2016_Maccio_Udresco_Dutton}, the Tully-Fisher relation \citep{2017_Dutton_Obreja_Wang}, the rotation curves of dwarf galaxies \citep{2018_SantosSantos_DiCintio_Brook} and the stellar mass versus black hole mass relation \citep{2019_Blank_Maccio_Dutton}.
Strictly speaking the results presented in this paper pertain to the NIHAO galaxies that we use for our analysis. However, in the previous 25 publications of the NIHAO series we have demonstrated that our simulations produce realistic galaxies, thus we are confident that the conclusions of this paper might also be applied to real galaxies.

We calculate the SFR at a specific time as the mass of all star particles within 20 per cent of the galaxy's virial radius that have formed within the last 200 Myr, divided by 200 Myr.
Some galaxies have simulation outputs with a SFR of zero. In that case we estimate a lower limit of the SFR as the gas particle mass divided by 200 Myr, and show them in our figures as arrows.
Our results are robust amongst different time intervals for calculating the SFR. Measuring with a shorter time interval merely increases the incidence of SFRs with value zero.

For Section \ref{sec:results1} we use 78 NIHAO galaxies from \citet{2015_Wang_Dutton_Stinson} and for Section \ref{sec:results2} a subset thereof containing 50 galaxies for which we have a dark matter counterpart.
Black holes were introduced into NIHAO by \citet{2019_Blank_Maccio_Dutton}, but 
only for Fig. \ref{fig:sfr_mstar_params} in Section \ref{sec:results1} we use 52 of these galaxies to demonstrate that AGN feedback significantly influences the slope of the SFMS.
Otherwise we only
use the original NIHAO galaxies from \citet{2015_Wang_Dutton_Stinson} that do not include black holes, for the following reasons:
only galaxies that are actually star forming are part of the SFMS, whereas quenched galaxies form a distinct population in the stellar mass versus SFR plane.
This usually poses great difficulties for observations and most simulations alike, as schemes have to be developed to filter quenched galaxies from the sample in order to be able to investigate the SFMS.
Furthermore even galaxies that are still within the SFMS can already be significantly affected by AGN feedback and thus distort the view on the SFMS, which we will show in Section \ref{sec:results1}.
The NIHAO sample of galaxies without black holes provide an opportunity to investigate the SFMS unaffected by AGN feedback and unbiased by selection effects and uncertainties due to the exclusion criteria of quenched galaxies.
How black holes influence the SFR of galaxies and how AGN feedback quenches star formation will be investigated in a separate publication.

\section{The Star Formation Main Sequence}\label{sec:results1}

\begin{figure*}
  \includegraphics[width=174mm]{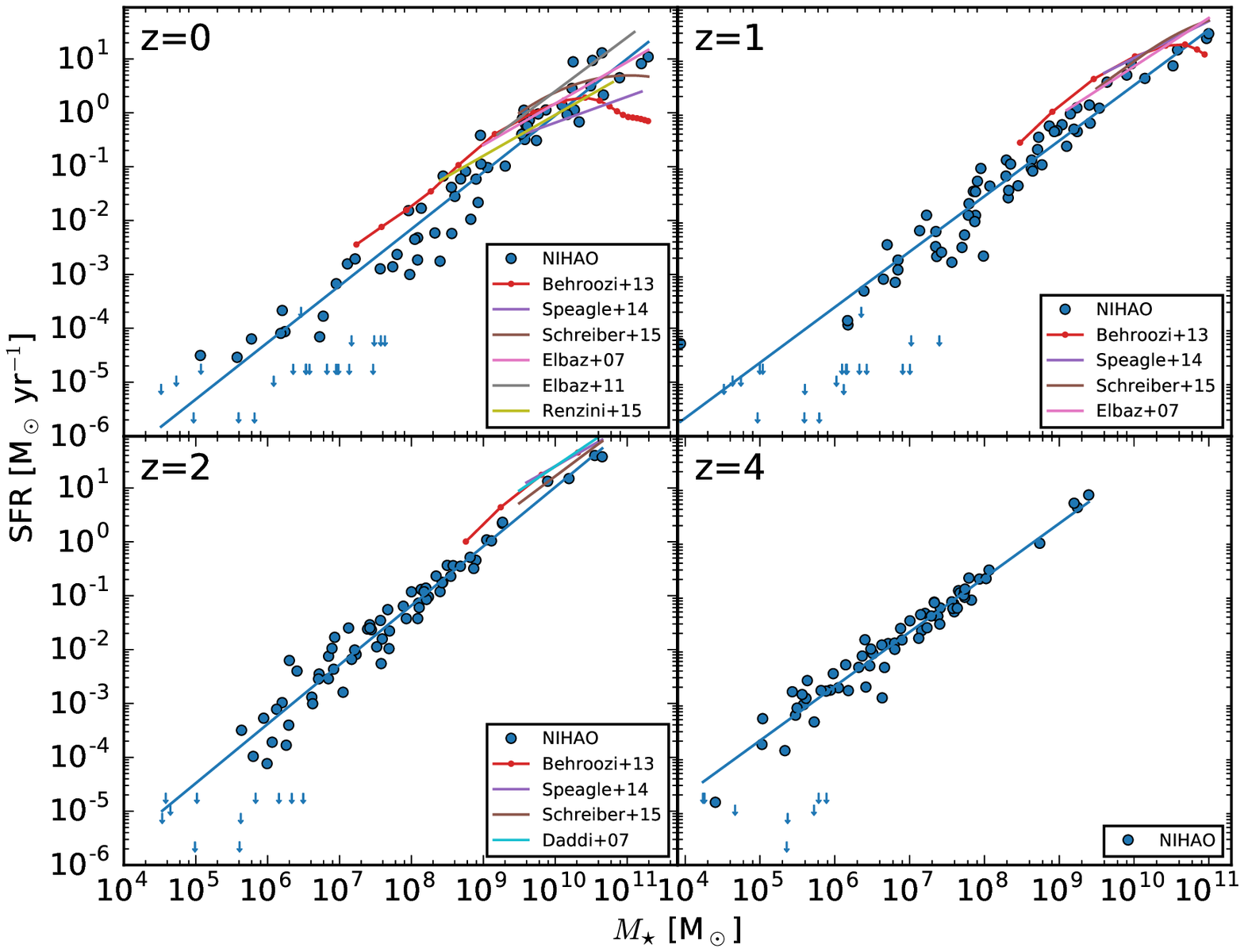}
  \caption{Star formation rate SFR versus stellar mass $M_{\star}$ for the NIHAO galaxies with a linear fit for four different redshifts, compared to several observations \citep{2013a_Behroozi_Wechsler_Conroy,2013b_Behroozi_Wechsler_Conroy,2014_Speagle_Steinhardt_Capak, 2015_Schreiber_Pannella_Elbaz, 2007_Elbaz_Daddi_LeBorgne, 2011_Elbaz_Dickinson_Hwang,2015_Renzini_Peng, 2007_Daddi_Dickinson_Morrison}.
  The arrows are a lower-limit estimate for the SFR for cases where the SFR is zero.
                }
  \label{fig:sfr_mstar_zx4}
\end{figure*}

In this section we investigate the SFMS of the NIHAO galaxies, its slope, normalization, scatter and evolution, and compare our findings to several observations.
Fig. \ref{fig:sfr_mstar_zx4} shows the relation between SFR and stellar mass of the NIHAO galaxies for four different redshifts.
At all redshifts (i.e. for all 64 simulation outputs) we fit a linear relation to $\log$ SFR and $\log M_{\star}$, which provides a good match to the simulation data.
The arrows denote a lower-limit estimate for the SFR for cases where the SFR is zero, we do not use these values for calculating the linear fit.
We also compare our results to several observations.
However, most observations are only available for large stellar masses of $\gtrsim 10^{9}\,\mathrm{M}_{\odot}$, whereas the NIHAO simulations cover stellar masses down to $\sim 10^{4}\,\mathrm{M}_{\odot}$.
\citet{2013a_Behroozi_Wechsler_Conroy,2013b_Behroozi_Wechsler_Conroy} provide a compilation of various observations from different publications.
At $z=0,1,2$ this data fits the NIHAO simulations quite well for stellar masses below $\sim 10^{11}\,\mathrm{M}_{\odot}$. At higher stellar masses the observed values drop to lower SFRs as star formation in these galaxies gets quenched by AGN feedback. This does not happen for the NIHAO galaxies as these do not include black holes.
Only the normalization of our fit for stellar masses below $\sim 10^{9}\,\mathrm{M}_{\odot}$ is slightly lower than the \citet{2013a_Behroozi_Wechsler_Conroy,2013b_Behroozi_Wechsler_Conroy} data.
Also other observations \citep{2014_Speagle_Steinhardt_Capak, 2015_Schreiber_Pannella_Elbaz, 2007_Elbaz_Daddi_LeBorgne, 2011_Elbaz_Dickinson_Hwang,2015_Renzini_Peng, 2007_Daddi_Dickinson_Morrison} coincide with the NIHAO simulations.

\begin{figure}
  \includegraphics[width=84mm]{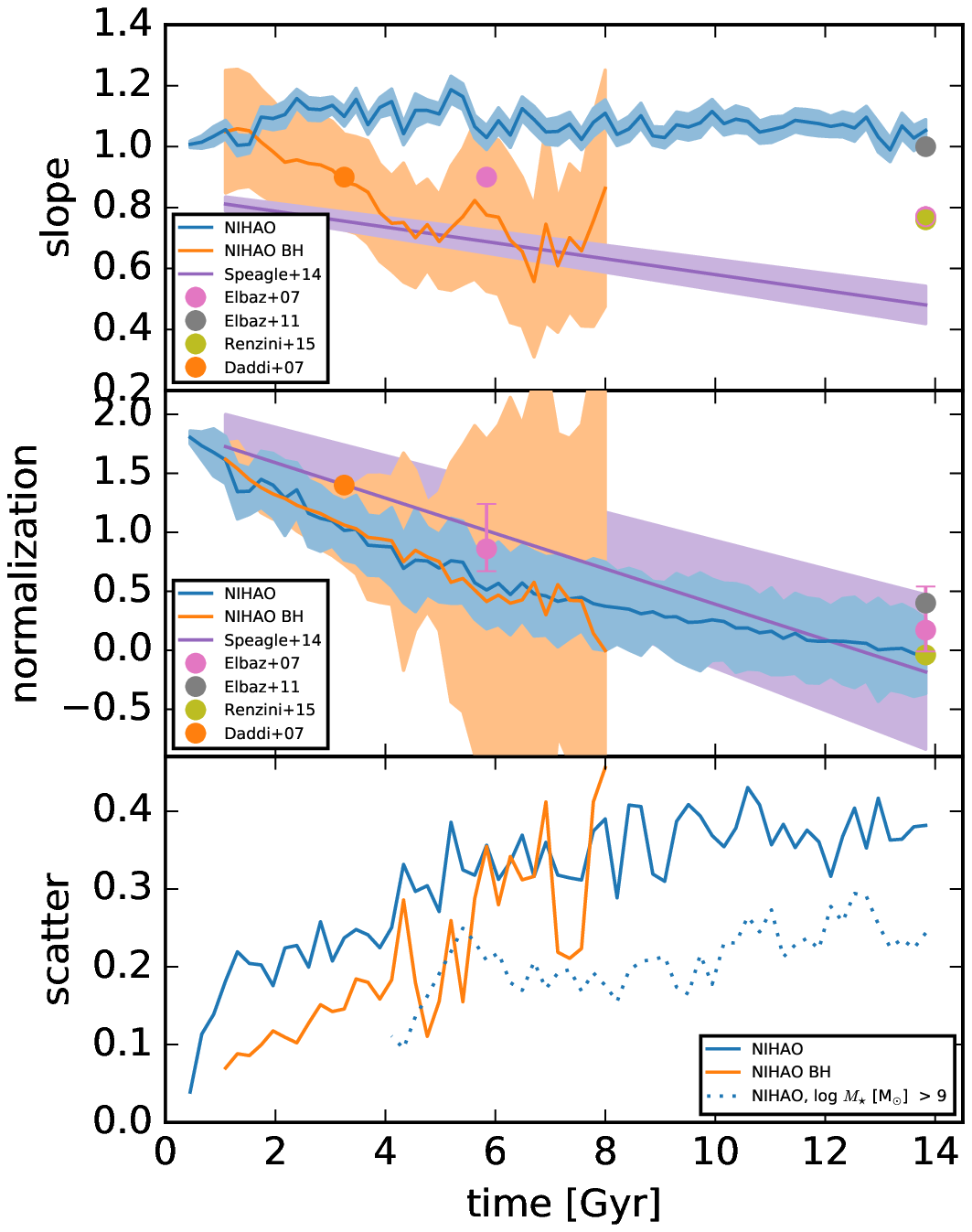}
  \caption{Slope, normalization (SFR at $10^{10}\,\mathrm{M}_{\odot}$) and scatter of the SFMS versus time for the NIHAO galaxies (with and without black holes), compared to several observations \citep{2014_Speagle_Steinhardt_Capak, 2007_Elbaz_Daddi_LeBorgne, 2011_Elbaz_Dickinson_Hwang,2015_Renzini_Peng, 2007_Daddi_Dickinson_Morrison}.
  In addition we show in the lower panel the scatter for NIHAO galaxies with a stellar mass $> 10^{9}\,\mathrm{M}_{\odot}$.
                }
  \label{fig:sfr_mstar_params}
\end{figure}

Fig. \ref{fig:sfr_mstar_params} shows the parameters of the linear fit, i.e. the slope, normalization (the SFR at stellar mass
$10^{10}\,\mathrm{M}_{\odot}$) and the scatter, compared to several observed values: \citet{2014_Speagle_Steinhardt_Capak, 2007_Elbaz_Daddi_LeBorgne, 2011_Elbaz_Dickinson_Hwang,2015_Renzini_Peng, 2007_Daddi_Dickinson_Morrison}.
Note that \citet{2013a_Behroozi_Wechsler_Conroy,2013b_Behroozi_Wechsler_Conroy} and \citet{2015_Schreiber_Pannella_Elbaz} do not fit a linear relation to their data, thus we omit them here.
Our slope is not evolving with time and has values of about unity, and thus is consistently higher than observed values, which range from about 0.5 to 1.
This can be explained as follows:
most observations cover quite high stellar masses of $\gtrsim 10^{9}\,\mathrm{M}_{\odot}$, where AGN feedback already starts to reduce the SFR of these galaxies, thus their relations are \lq bending down,\rq \, yielding a smaller slope.
Thus the \lq natural,\rq \lq zero-order\rq \, slope of the SFMS is around unity, which is then reduced to lower values at high stellar masses due to the influence of AGN feedback.
\citet{2014_Abramson_Kelson_Dressler} also report a declining slope with increasing stellar mass, which they attribute to mass quenching due to the build-up of a stellar bulge.

\begin{figure}
  \includegraphics[width=84mm]{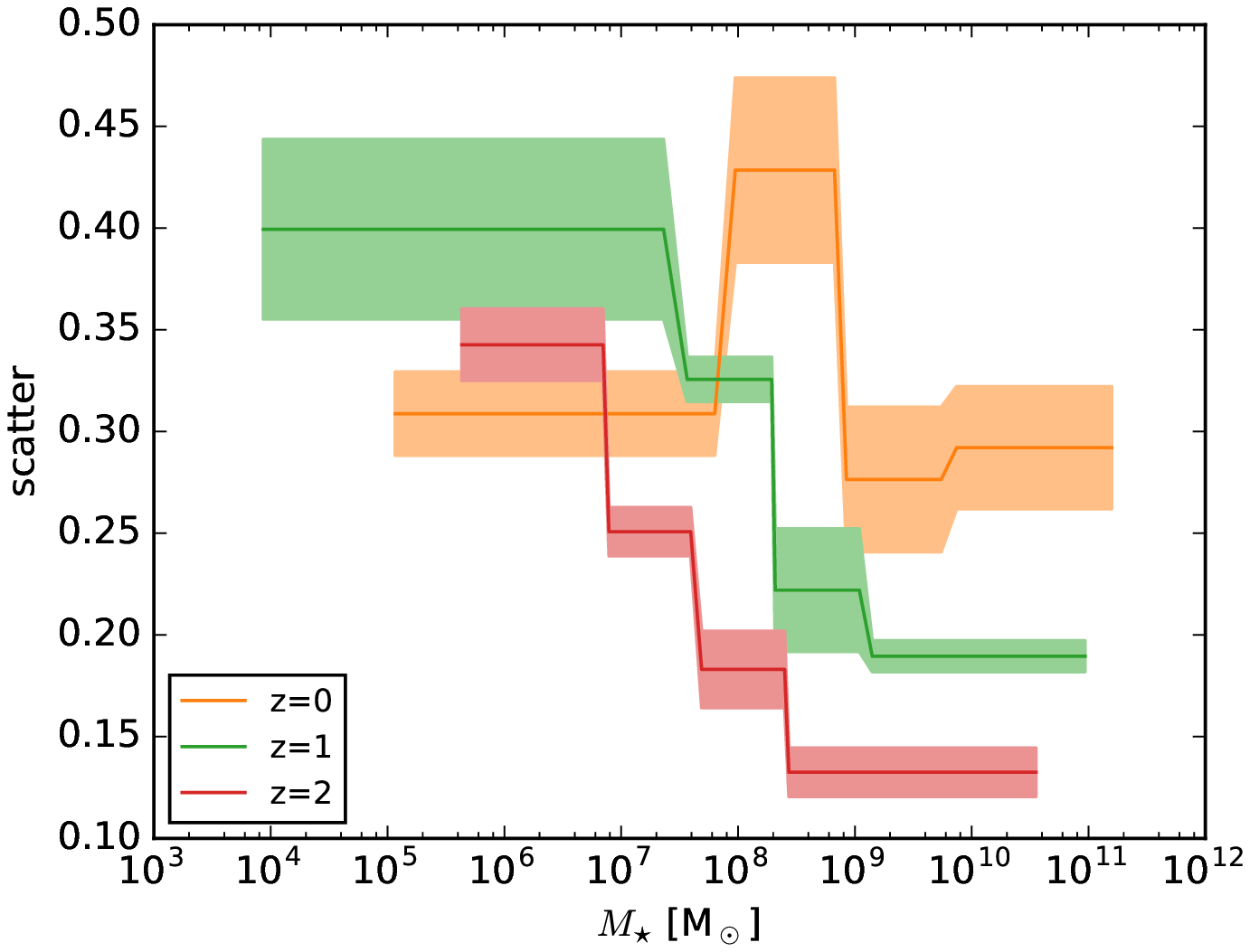}
  \caption{Scatter of the SFMS as a function of stellar mass for redshifts 0,1,2.
  Each bin of stellar mass contains at least 10 galaxies.
  The variance is estimated with the jacknife resampling method.
                }
  \label{fig:scatter_mstar}
\end{figure}

The normalization in the middle panel of Fig. \ref{fig:sfr_mstar_params} shows a good agreement of the NIHAO simulations with several observed values.
The lower panel of Fig. \ref{fig:sfr_mstar_params} shows the scatter of the SFMS.
Observations usually report a scatter of about 0.2-0.3 dex, that does not depend on redshift or stellar mass. 
However, the scatter of the NIHAO galaxies is around 0.35 for up to redshift one, and thus  higher than observed values of 0.2-0.3.
This is because in the NIHAO simulations lower stellar masses ($\lesssim 10^{9}\,\mathrm{M}_{\odot}$), which are usually not covered in most observations, yield a higher scatter than higher stellar masses, which then contributes to the overall scatter shown in Fig. \ref{fig:sfr_mstar_params}.
To demonstrate this effect we recalculate the scatter by using only galaxies with stellar masses $> 10^{9}\,\mathrm{M}_{\odot}$, shown in the lower panel of Fig. \ref{fig:sfr_mstar_params} as dotted line. We again omit redshifts with less than 10 galaxies. Indeed the scatter is now much lower with about 0.25 dex and thus in agreement with observed values.
Additionally we show in Fig. \ref{fig:scatter_mstar} the scatter of the SFMS as a function of stellar mass for three different redshifts. Each bin of stellar mass contains at least 10 galaxies.
As the number of galaxies per bin is low, we estimate the variance of the scatter with the jackknife resampling method. For stellar masses $\gtrsim 10^{8}\,\mathrm{M}_{\odot}$  the scatter is increasing with decreasing stellar mass, and decreasing with redshift.
We note that for very low stellar masses ($\lesssim 10^{7}\,\mathrm{M}_{\odot}$) the scatter might be artificially increased due to the stochasticity of star formation, as some galaxies only form few star particles.

To illustrate how AGN feedback reduces the slope of the SFMS we take 52 NIHAO galaxies from \citet{2019_Blank_Maccio_Dutton} which have been simulated with black holes, and calculate the slope, intercept and scatter of the SFMS for these galaxies. We exclude completely quenched galaxies by only selecting those whose SFR is less than three times the scatter away from the fitted SFMS of Fig. \ref{fig:sfr_mstar_zx4}.
We also exclude those with a stellar mass $< 10^{9}\,\mathrm{M}_{\odot}$
which are usually not covered in most observations.
We furthermore only calculate these quantities when there are at least 10 galaxies fulfilling these criteria, therefore there is no data for low redshifts where most galaxies are quenched. 
This slope is shown in the upper panel Fig. \ref{fig:sfr_mstar_params}, it is indeed significantly lower than the slope of around unity for the NIHAO galaxies without black holes, and fits well with the observed slope of \citet{2014_Speagle_Steinhardt_Capak}.
The intercept does not change significantly, only its standard deviation grows towards smaller redshifts.
The galaxies deviate from the SFMS due to AGN feedback, leading to a less tight relation.
Also the scatter is at or below the scatter for NIHAO galaxies without black holes,
showing that the scatter is not dominated by the effects of AGN feedback.

\section{The origin of the scatter of the Star Formation Main Sequence}\label{sec:results2}

\begin{figure}
  \includegraphics[width=84mm]{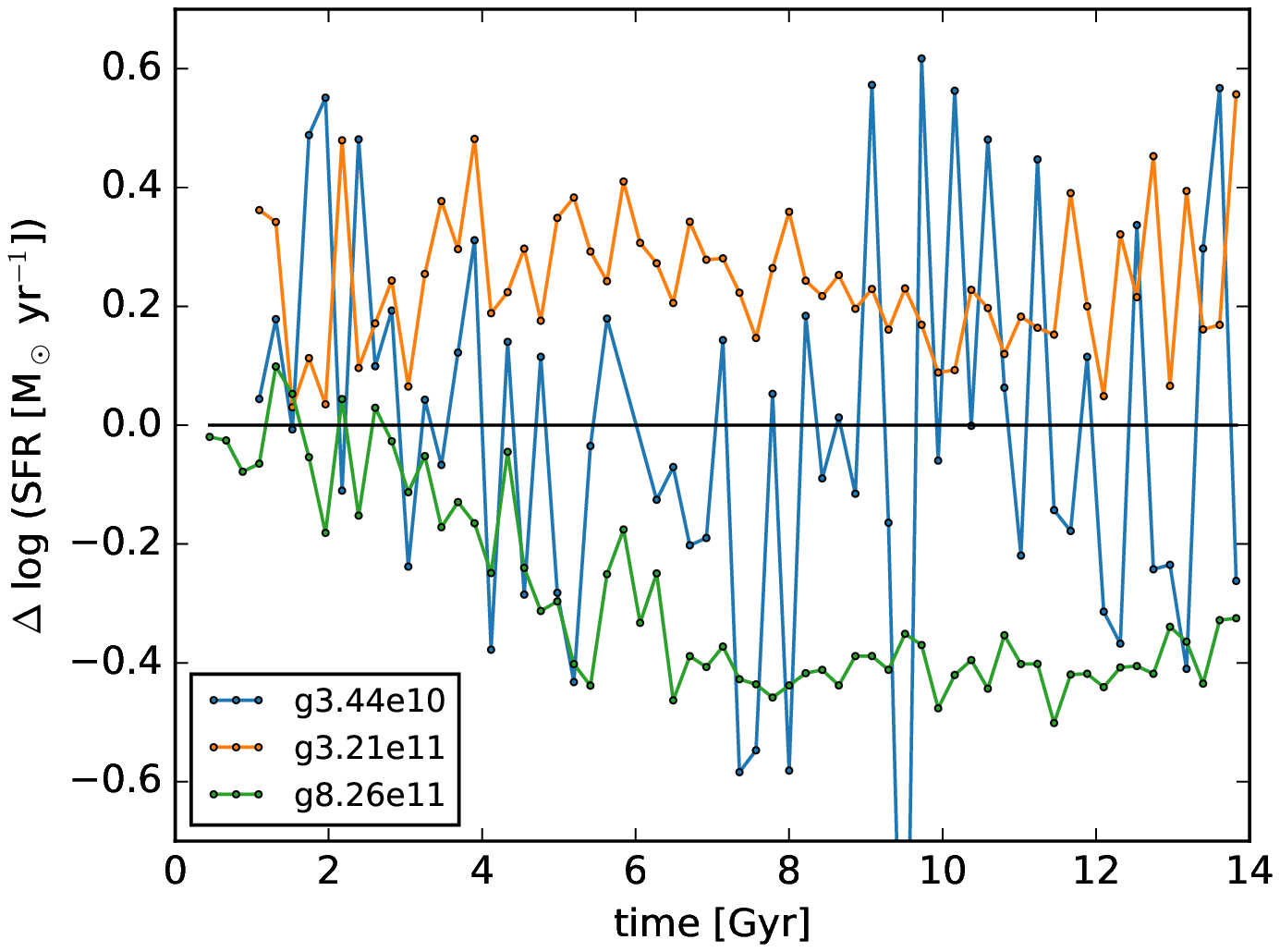}
  \caption{$\Delta \log \mathrm{SFR}$ versus time for three of the NIHAO galaxies.
                }
  \label{fig:time_dlogsfr}
\end{figure}

In this section we investigate why and how galaxies deviate from the SFMS, and what is influencing its scatter.
Our fits to the SFMS (Fig. \ref{fig:sfr_mstar_zx4}) give us an average star formation rate SFR$_\mathrm{avg}$ as a function of redshift and stellar mass.
Therefore we define the deviation from the average SFR as
\begin{equation}
\Delta \log \mathrm{SFR} = \log \mathrm{SFR} - \log \mathrm{SFR}_\mathrm{avg} \,.
\end{equation}
Thus an \lq average\rq \, galaxy would have a value of $\Delta \log \mathrm{SFR} = 0$ for all times,
or fluctuate slightly around zero.
However, many galaxies deviate from this expected behavior as shown in Fig. \ref{fig:time_dlogsfr}, which shows $\Delta \log \mathrm{SFR}$ versus time for three galaxies.
The galaxy g3.44e10\footnote{The name of each galaxy indicates
approximately its halo mass at $z=0$.} (blue line) indeed fluctuates around zero throughout its lifetime, whereas galaxy g3.21e11 (orange line) is a chronic overproducer of stars, and galaxy g8.26e11 (green line) is a chronic underproducer.
What is driving the different behavior of these galaxies?
According to \citet{2010_Dutton_Bosch_Frank} the scatter of the SFMS is dependent on the halo concentration, therefore we investigate the correlation of $\Delta \log \mathrm{SFR}$ with the galaxies' $z=0$ NFW \citep{1996_Navarro_Frenk_White} halo concentration.
We use the halo concentration of the corresponding dark-matter-only (DMO) simulations, as baryonic effects can lead to non-NFW dark matter profiles \citep[e.g.,][]{2020_Maccio_Crespi_Blank}. From the 78 galaxies used in the previous section 50 have a DMO counterpart.

\begin{figure}
  \includegraphics[width=84mm]{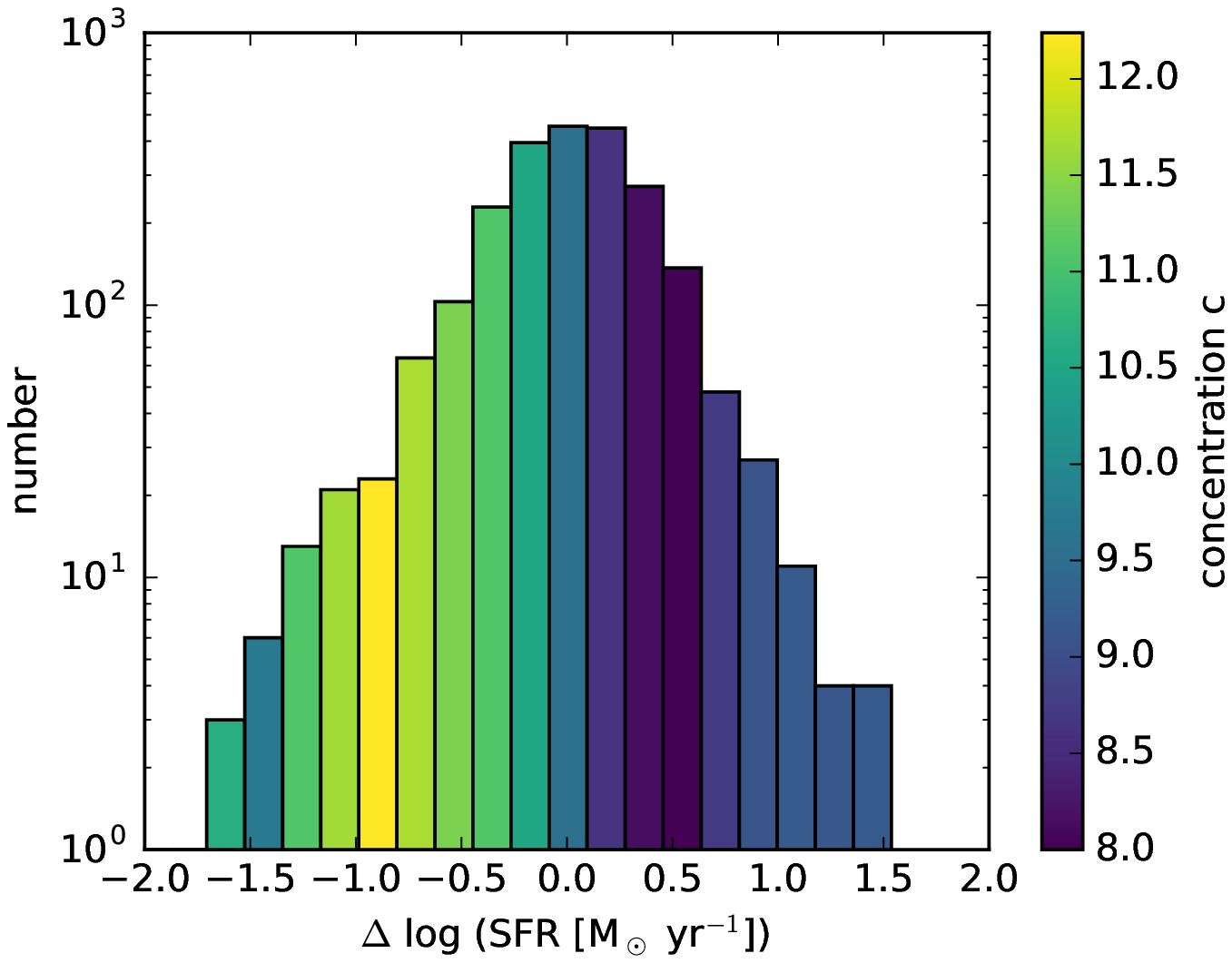}
  \caption{Number of galaxies in each bin of $\Delta \log \mathrm{SFR}$, color coded with the average $z=0$ halo concentration.
                }
  \label{fig:hist}
\end{figure}

\begin{figure}
  \includegraphics[width=84mm]{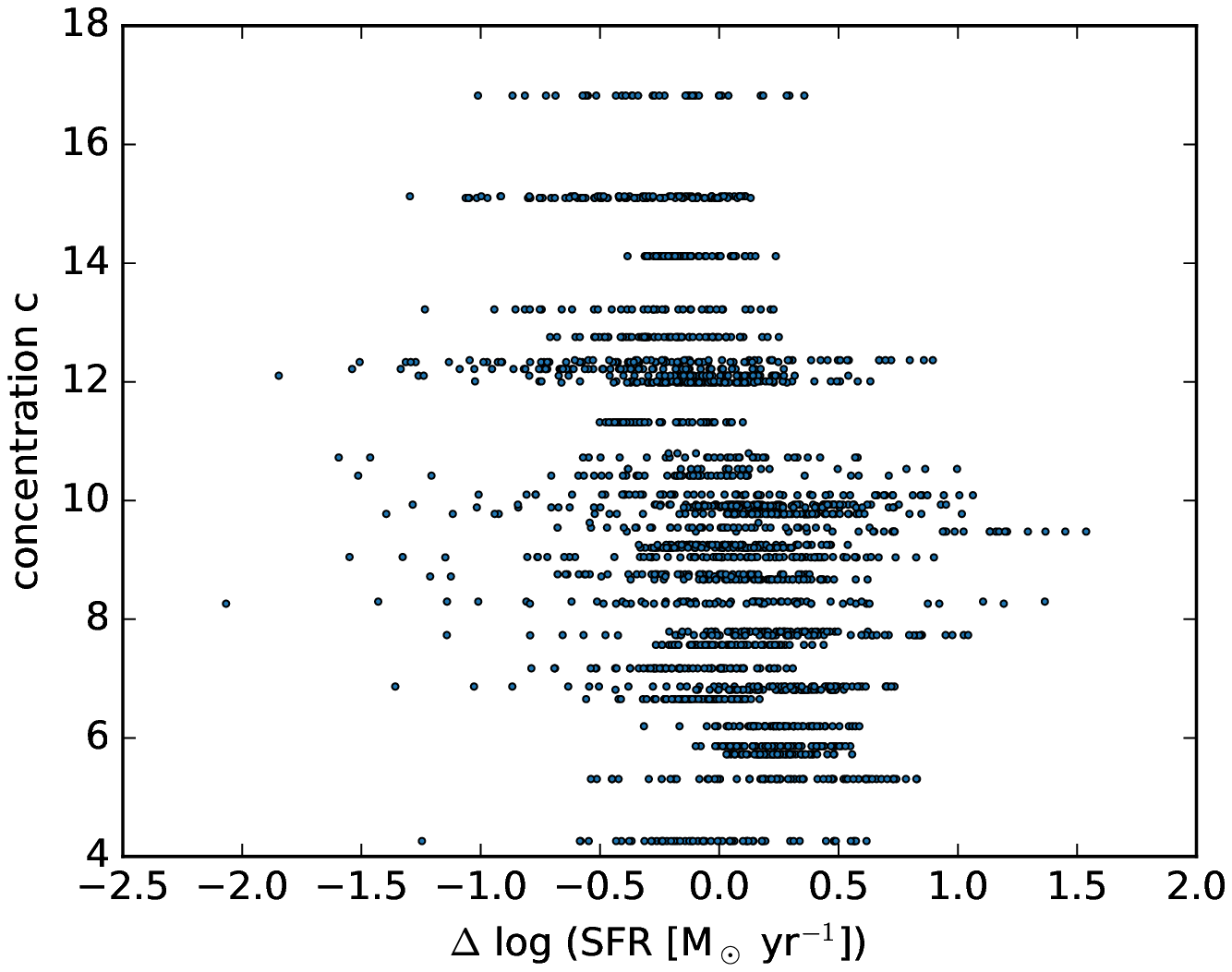}
  \caption{Relation between $z=0$ halo concentration and $\Delta \log \mathrm{SFR}$.
                }
  \label{fig:dsfr_cc}
\end{figure}

The histogram in Fig. \ref{fig:hist} shows the number of galaxies in each bin of $\Delta \log \mathrm{SFR}$, color coded with the average $z=0$ halo concentration of each bin. Here we use all 50 galaxies at all 64 snapshots, except those that have a SFR of zero.
\begin{figure}
  \includegraphics[width=84mm]{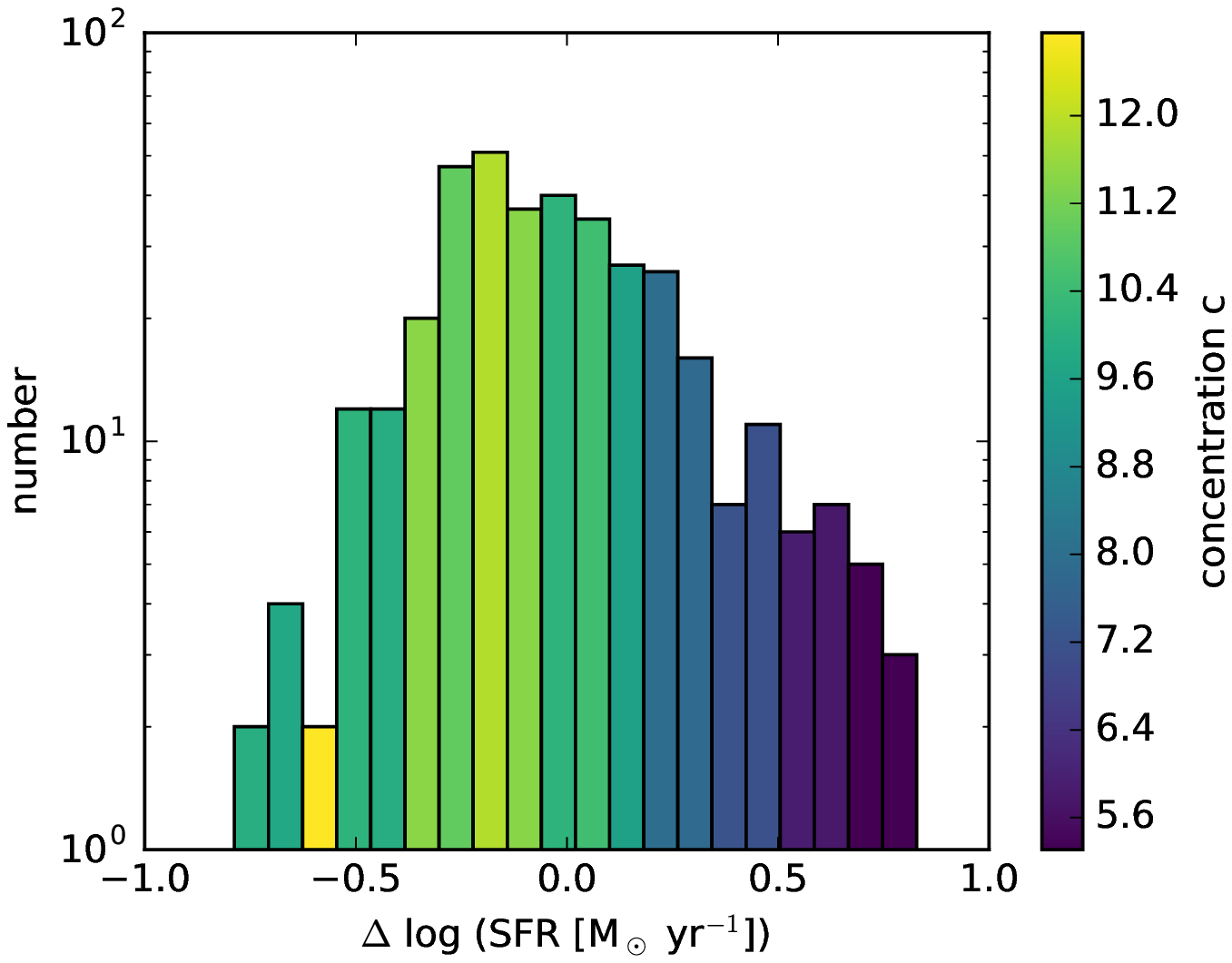}
  \caption{Number of galaxies in each bin of $\Delta \log \mathrm{SFR}$, color coded with the average $z=0$ halo concentration for six galaxies with halo mass range of $1-2 \times 10^{11} \mathrm{M}_{\sun}$.
                }
  \label{fig:hist_slct}
\end{figure}
Chronic overproducers, i.e. galaxies with $\Delta \log \mathrm{SFR} > 0$, tend to have lower $z=0$ halo concentrations, whereas chronic underproducers ($\Delta \log \mathrm{SFR} < 0$) tend to have higher $z=0$ halo concentrations.
Fig. \ref{fig:hist_slct} shows a similar histogram, but only for six galaxies with halo mass
range of 1-2$\times 10^{11}\, \mathrm{M}_{\sun}$. This shows that the effect shown in Fig. \ref{fig:hist} is not caused by variations in galaxy mass.
Fig. \ref{fig:dsfr_cc} furthermore shows the $z=0$ halo concentration versus $\Delta \log \mathrm{SFR}$ for all times.
A Spearman correlation gives a p-value of 0.014, clearly showing the existence of a correlation between these two quantities.

\begin{figure}
  \includegraphics[width=84mm]{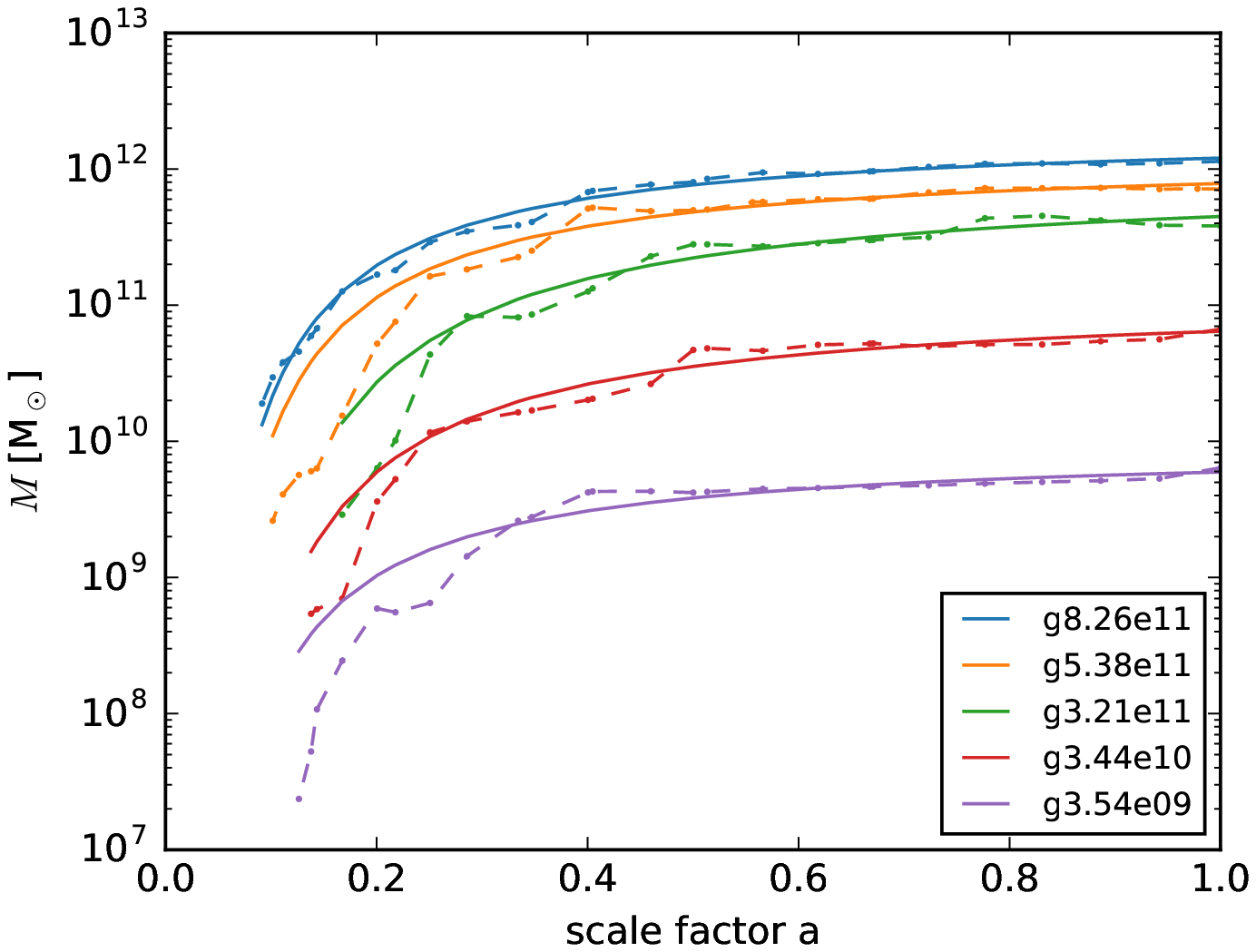}
  \caption{Halo mass as function of scale factor for five galaxies.
               Dotted lines: simulation data, solid lines: fit according to equation (\ref{eq:mhalogrow}).
                }
  \label{fig:A_m_a_dmo}
\end{figure}

We next investigate why the $z=0$ halo concentration is correlated with the SFR across cosmic time. The concentration is dependent on the cosmic matter density at which the halo forms, thus the concentration is also correlated with the halo formation time.
We follow \citet{2002_Wechsler_Bullock_Primack} to calculate the halo formation time (i.e. the scale factor at which the halo forms). The mass growth of a halo as a function
of the scale factor $a$ can be described as
\begin{equation}
  M(a) = M_{0} \exp \left[ - 2 a_{\mathrm{c}} \left(\frac{1}{a}-1\right) \right]
  \label{eq:mhalogrow}
\end{equation}
where $M_{0}$ is the $z=0$ halo mass, and $a_{\mathrm{c}}$ can be interpreted as the scale factor at which the halo forms.
We fit equation (\ref{eq:mhalogrow}) to the halo growth histories of
the DMO simulations. In Fig. \ref{fig:A_m_a_dmo} we show the halo growth histories and their fits for five galaxies, showing that equation (\ref{eq:mhalogrow}) provides a reasonable description for the halo growth.

\begin{figure}
  \includegraphics[width=84mm]{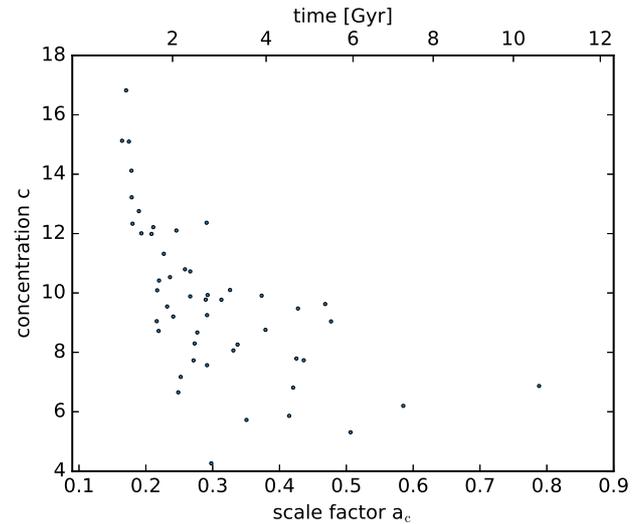}
  \caption{Correlation between $z=0$ halo concentration and formation scale factor $a_{\mathrm{c}}$.
                }
  \label{fig:A_cc_ac}
\end{figure}

\begin{figure}
  \includegraphics[width=84mm]{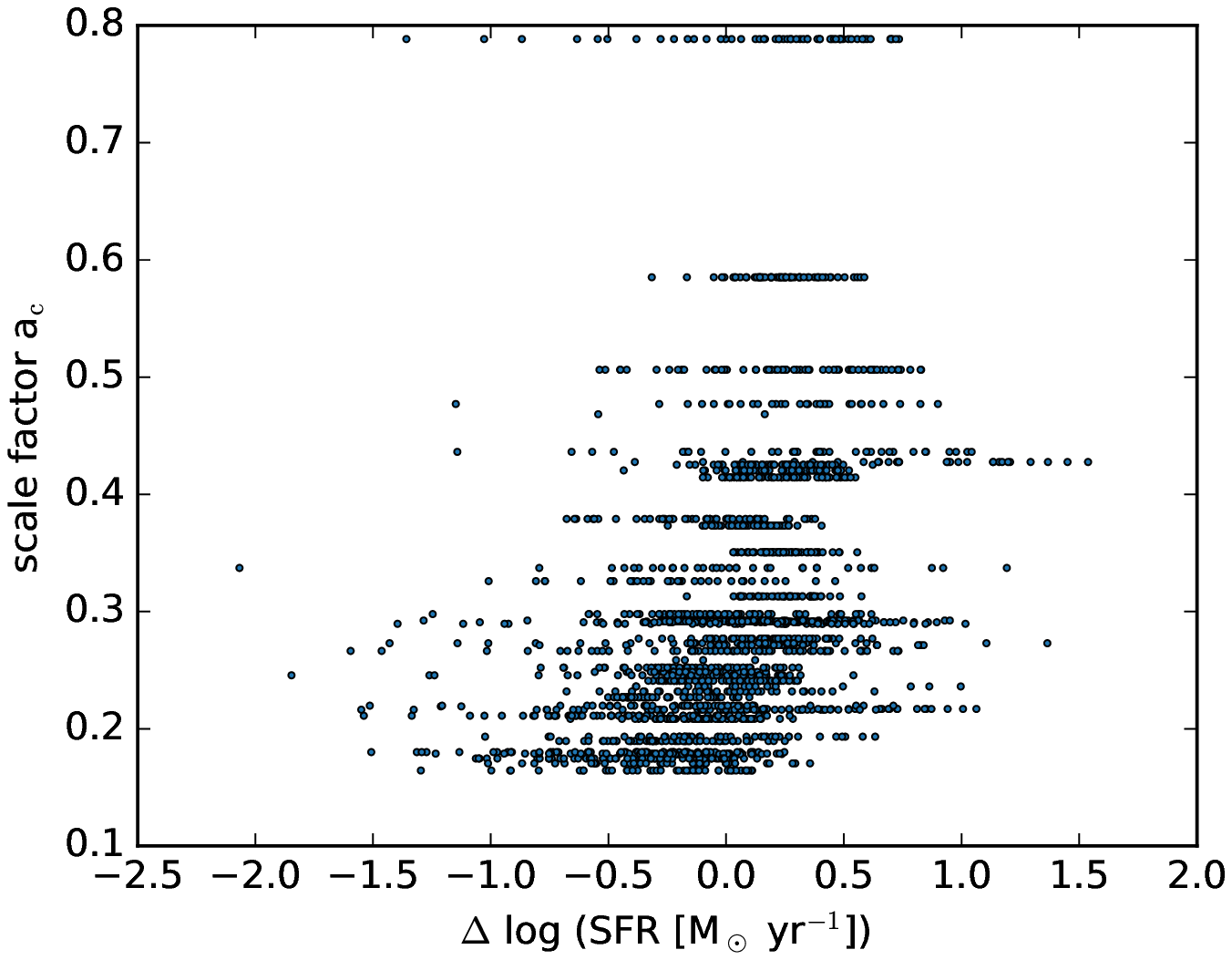}
  \caption{Correlation between halo formation scale factor $a_{\mathrm{c}}$ and $\Delta \log \mathrm{SFR}$.
                }
  \label{fig:A_dsfr_ac}
\end{figure}

The fitting procedure
gives us the formation time, i.e. the formation scale factor $a_{\mathrm{c}}$ for each galaxy.
We then show the correlation between $z=0$ halo concentration and formation scale factor $a_{\mathrm{c}}$ in Fig. \ref{fig:A_cc_ac}, and the correlation between halo formation scale factor $a_{\mathrm{c}}$ and $\Delta \log \mathrm{SFR}$ in Fig. \ref{fig:A_dsfr_ac}.
With a p-value of 0.002 this correlation is stronger than the correlation between $\Delta \log \mathrm{SFR}$ and the $z=0$ halo concentration.
Thus haloes that form later, i.e. have a lower halo concentration, have a higher SFR and
haloes that form earlier, i.e. have a higher halo concentration, have a lower SFR than an average galaxy.
This means that if a galaxy lives above or below the SFMS is set at birth, which can be explained as follows:
Galaxies that are overproducers of stars seem to have a higher-than-average SFR, they are shifted above the SFMS.
However, a perhaps better
interpretation is that they are shifted to the left from the SFMS, thus
{\it they do not have a higher-than-average SFR, but a lower-than-average stellar mass}
(with respect to the SFMS of the same SFR).\footnote{Although galaxies above the SFMS grow faster (in stellar mass) than galaxies below it, their tracks in the SFR versus $M_{\star}$ plane do not necessarily cross, as the ratio of SFR and $M_{\star}$, i.e. their slope, is always around unity. However, their tracks can cross in the $M_{\star}$ versus time plane.}
This corresponds to a later formation time, because these galaxies did not have enough time yet to form sufficient stars. Likewise underproducers of stars do not have a lower-than-average SFR, but a higher-than-average stellar mass, because they have formed stars for a longer time.
It is of course to some extent arbitrary if the deviation of galaxies from the SFMS is measured in $M_{\star}$ or SFR direction. However, choosing $M_{\star}$ offers a physically more plausible interpretation than choosing the SFR.
Finally, according to \citet{2020_Wang_Lilli_A,2020_Wang_Lilli_B} fluctuations on time-scales < 200 Myr might also contribute to the the scatter of the SFMS; therefore, the effects we describe in this paper might not be the sole cause of the scatter.

\begin{figure}
  \includegraphics[width=84mm]{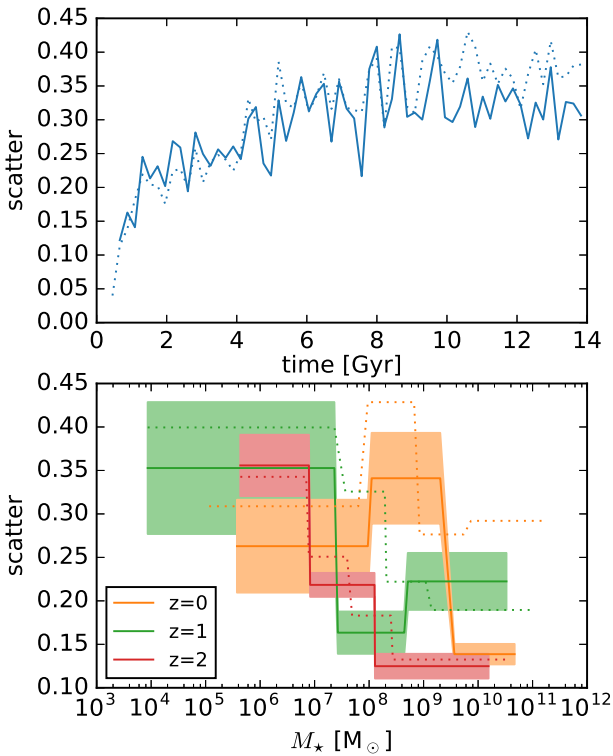}
  \caption{Upper panel: original scatter of the NIHAO galaxies (dotted line, same as in Fig. \ref{fig:sfr_mstar_params}) compared to the scatter of the \lq corrected\rq \, SFR (solid line, according to equation (\ref{eq:sfrcorr})) as a function of redshift.
Lower panel: original scatter of the NIHAO galaxies (dotted lines, same as in Fig. \ref{fig:scatter_mstar}) compared to the scatter of the \lq corrected\rq \, SFR (solid lines, according to equation (\ref{eq:sfrcorr})) as a function of stellar mass for three redshifts.
For the latter we use fewer galaxies (only the ones with DMO counterpart), thus we divide them into only three bins. The variance is estimated with the jacknife resampling method.
                }
  \label{fig:scatter_corr}
\end{figure}

Thus the halo formation time influences the scatter of the SFMS.
Following \citet{2019_Matthee_Schaye} we fit a linear relation to $\Delta \log \mathrm{SFR}$ as a function of halo formation scale factor, thus we can calculate a \lq corrected\rq \, or \lq residual\rq \, SFR as
\begin{equation}
  \log \mathrm{SFR}_{\mathrm{corr}} = \log \mathrm{SFR} + \alpha + \beta a_{\mathrm{c}} \,,
  \label{eq:sfrcorr}
\end{equation}
where the SFRs are measured in $\mathrm{M}_{\odot} \, \mathrm{yr}^{-1}$.
The upper panel of Fig. \ref{fig:scatter_corr} shows the scatter of this corrected SFR compared to the original scatter from Fig. \ref{fig:sfr_mstar_params}. The overall scatter is reduced by about 0.05 dex, in accordance with \citet{2019_Matthee_Schaye}.
However, the reduction in the scatter only reaches as far back as 5 Gyr. For earlier times both, the corrected SFR and the original SFR, yield about the same scatter.
As a function of stellar mass (lower panel of Fig. \ref{fig:scatter_corr}) the scatter of the corrected SFR is lower only for $z=0$ down to $\sim 10^8\,\mathrm{M}_{\odot}$. For $z=1$ and $z=2$ no reduction in the scatter occurs, in agreement with the upper panel of Fig. \ref{fig:scatter_corr}.

\section{Summary}\label{sec:summary}
In this paper we investigate the SFMS and the origin of its scatter by using the NIHAO suite of galaxy simulations.
In the first part we compare the SFMS of the NIHAO galaxies with several observations. They generally agree with each other, but the slope of the SFMS is around unity for all redshifts and thus significantly higher than values derived from observations.
This is because galaxies at high stellar masses ($\gtrsim 10^{9}\,\mathrm{M}_{\odot}$) are already influenced by AGN feedback, thus their tracks in the stellar mass versus SFR plane are not linear but \lq bending down,\rq \, leading to lower observed slopes.
The NIHAO galaxies without AGN feedback show that the \lq zero-order\rq \, slope of the SFMS, unaffected by AGN feedback, is around unity.
We confirm this by recalculating the SFMS with 52  NIHAO galaxies with black holes from \citet{2019_Blank_Maccio_Dutton}, whose slope is significantly smaller than unity and in agreement with observed slopes.
Most observations report a scatter of 0.2-0.3 dex \citep[see e.g.][]{2019_Donnari_Pillepich_Nelson} independent of stellar mass, but the NIHAO simulations show that the scatter is
only that small at large stellar masses. As we find that small stellar masses ($\lesssim 10^{9}\,\mathrm{M}_{\odot}$) have larger scatter, we report an overall total scatter of 0.35 dex.
Recalculating the scatter by using only galaxies with $M_{\star} > 10^{9}\,\mathrm{M}_{\odot}$ gives a scatter of 0.25 which is in agreement with observed values.

In the second part of the paper we investigate the origin of the scatter of the SFMS.
We calculate the deviation from the fitted SFMS for each galaxy. This deviation does not fluctuate around zero as would be expected, but is either above or below zero for most of the evolution, i.e. these galaxies are either chronic over- or underproducers of stars.
We find that the SFMS deviation correlates with the galaxy's dark matter halo concentration at $z=0$ \citep[see also][]{2010_Dutton_Bosch_Frank}, with overproducers having a low halo concentration and vice versa. 
We also confirm that halo concentration is anticorrelated with halo formation time \citep[see also][]{2002_Wechsler_Bullock_Primack}. Thus a galaxy's deviation from the SFMS is correlated with the halo formation time \citep[see also][]{2019_Matthee_Schaye}: overproducers of stars form later and underproducers form earlier.
What is the origin of this behavior?
Does \enquote{a galaxy's SFR remembers its past SFR} \citep{2019_Matthee_Schaye}?
This is not the case. We provide the following simple explanation for this phenomenon:
instead of interpreting these galaxies as high-SFR or low-SFR (as they lie above or below the SFMS) they can be interpreted as low-stellar-mass or high-stellar-mass galaxies (as they lie to the left or to the right of the SFMS). That means that later forming galaxies have a lower stellar mass than average, and vice versa. This is easy to comprehend: Later forming galaxies have less time to form stars, thus they have a lower-than-average stellar mass, and earlier forming galaxies have more time to form stars, thus they have a higher-than-average stellar mass.
It is therefore their {\it nature} (how and when these galaxies form), and not how they were {\it nurtured} (their local environment) that
sets the journey of a galaxy in the SFR versus stellar mass plane.

\section*{Acknowledgements}

The authors gratefully acknowledge the Gauss Centre for Supercomputing e.V. (www.gauss-centre.eu) for funding this project by providing computing time on the GCS Supercomputer SuperMUC at Leibniz Supercomputing Centre (www.lrz.de).
A part of this research was carried out on the High Performance Computing resources at New York University Abu Dhabi.
We used the software package {\sc pynbody} \citet{pynbody} for our analyses.

\section*{Data Availability Statement}

The data underlying this article will be shared on reasonable request to the corresponding author.

\bibliographystyle{mnras}
\bibliography{library}

\bsp
\label{lastpage}
\end{document}